\documentclass[a4paper,11pt]{article}
\usepackage{pos}

\title{Open heavy flavors: Theory}

\author*{Santosh K. Das}

\affiliation{School of Physical Sciences, Indian Institute of Technology Goa, Ponda-403401, Goa, India}

\emailAdd{santosh@iitgoa.ac.in}

\abstract{A brief overview of the theory of open heavy flavor dynamics in QCD matter produced in high energy heavy-ion collisions is presented. First, we will summarise the phenomenological efforts to estimate the  heavy quark diffusion coefficients obtained within different models. Then, the recent theoretical developments from different groups to probe the medium properties using heavy quarks will be presented. Heavy quarks are also considered as an ideal probe for the initial stage of heavy-ion collisions. In the end, we present the recent theoretical progress made to probe the early-stage effects, pre-equilibrium phase, and electromagnetic fields. }

\FullConference{%
 11th International Conference on Hard and Electromagnetic Probes of High-Energy Nuclear Collisions \\
 26-31 March 2023 \\
  Aschaffenburg, Germany
}

\begin{document}
\maketitle

\section{Introduction}
Heavy quarks (HQs)~\cite{Rapp:2018qla,Aarts:2016hap,Cao:2018ews} are considered an ideal probe of the quark-gluon plasma (QGP) phase produced in high-energy nuclear collisions. Due to their large masses ($M_c \sim 1.5$ GeV, $M_b\sim 4.5$ GeV), heavy quark-antiquark pairs are produced at the very early stage of high-energy heavy-ion collisions. As a outcome, heavy quarks witness the entire space-time evolution of the system and can act as an effective probe of the created matter. Furthermore, the thermalization time of heavy quarks is delayed relative to the light partons of the bulk medium by a factor of order $\sim M/T$, which renders it comparable to the lifetime of the QGP fireball. Thus, heavy quarks are not expected to fully thermalize and therefore preserve a memory of their interaction history, which can serve as a gauge of their interaction strength with the bulk medium. Being a non-equilibrium probe and produced at very early stage, they can also act as an excellent probe of the initial stage of heavy-ion collisions, i.e., initial electromagnetic field and pre-thermal phase. The study of heavy quark dynamics in QCD matter
requires considering its propagation through the pre-equilibrium, QGP, and hadronic phases after hadronization.

\section{Heavy quark diffusion coefficients}
One of the prime goals of all phenomenological studies~\cite{He:2012df,Song:2015ykw,Scardina:2017ipo,Xu:2017obm,Liu:2021dpm,Berrehrah:2014tva} of open heavy-flavor observables is to extract the heavy quark spatial diffusion coefficient, $D_x$, which quantify the interaction of heavy quarks with the bulk medium. $D_x$ is also directly related to the heavy quark thermalization time  and can be evaluated using lattice QCD (lQCD). Transport models are quite successful in describing the experimentally measured observables, the nuclear modification factor, $R_{AA}$,  and elliptic flow, $v_2$, of the D-meson.
Drag and momentum diffusion coefficients are inputs in the transport models (cross sections are the inputs in the Boltzmann equation), which contain the microscopic details mentioning how the heavy quarks interact with the hot QGP medium. Then, they compare their results with the experimental data. The value of the transport coefficients with which they can describe the 
experimental data, they relate it with the spatial diffusion coefficient, $D_x$.
The spatial  diffusion coefficient, $D_x$, 
can be calculated in the static limit ($p\rightarrow 0$) from the drag coefficient $D_x=T/M\Gamma$, where $T$ is the temperature of the thermal bath, $M$ is the mass of the heavy quark and $\Gamma$ is the drag coefficient. The standard quantification of the space
diffusion coefficient is done in terms of a dimensionless quantity $2\pi T D_x$ ~\cite{He:2012df} which is independent of mass.  One can estimate the charm quark thermalization time as $ \tau_{th} \equiv \Gamma^{-1}(p\rightarrow 0)$~\cite{Scardina:2017ipo}:

\begin{equation}
\tau_{th} = \frac{M D_x}{T}  = \frac{M}{2\pi T^2} (2\pi T D_x) \cong 1.8 \, \frac{2\pi T D_x}{(T/T_c)^2} \,\, \rm fm/c
\label{eq:th}
\end{equation}

\begin{figure*}
\centering
\includegraphics[clip,height=8 cm]{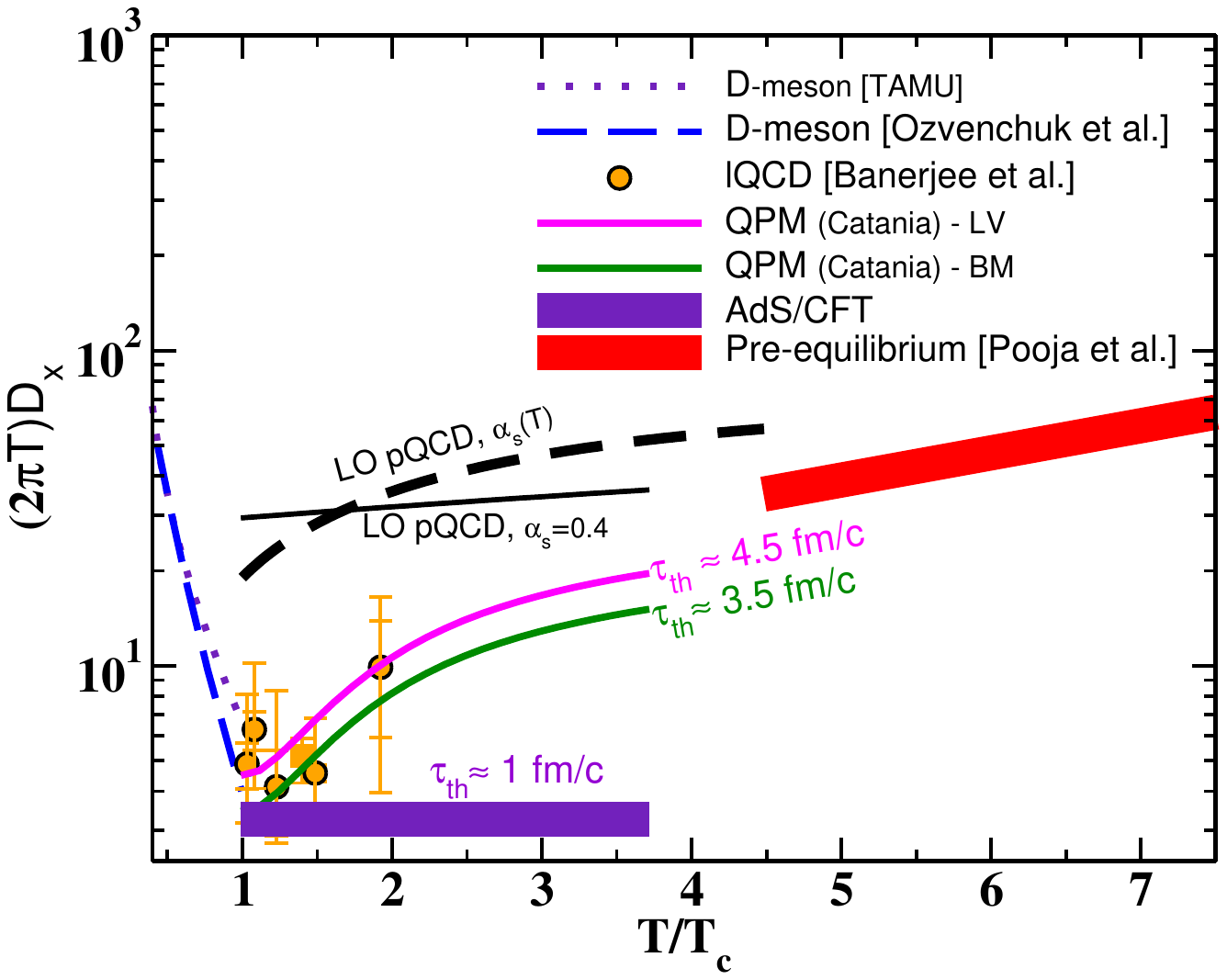}
\caption{ Spatial diffusion coefficient, $D_x$, as a function of temperature obtained within the Boltzmann transport approach (BM) and Langevin dynamics (LV) which can describe experimental data in comparison with the results from lattice QCD, along with $D_x$ of the D-meson in hadronic phase and c quark in pre-equilibrium phase (Glasma).} 
\label{fig:Dx}
\end{figure*}

In Fig.~\ref{fig:Dx}, the temperature variation of the $D_x$ for the charm quark in QGP and pre-equilibrium phase is shown along with $D_x$ of the D-meson in the hadronic phase.  $D_x$ of the charm quark in the QGP phase is obtained within the Boltzmann transport approach (BM) and Langevin dynamics (LV) which can describe experimental data, the $R_{AA}$ and $v_2$ of D-meson~\cite{Scardina:2017ipo}. 
$D_x$ of the charm quark in the pre-equilibrium phase, the Glasma phase, obtained within the framework Wong equation~\cite{Khowal:2021zoo}, is indicated in the same plot. In the pre-equilibrium phase, indeed, it is the effective temperature obtained from the energy density. First, the momentum diffusion coefficient in the expanding  Glasma phase is obtained considering $\sigma_p=2Dt$ at different $Q_s$, the  saturation scale. For the details, please refer to ref.~\cite{Khowal:2021zoo}. Then the effective temperature is obtained from the energy density at $t=0.5$ fm, considering the expansion, hence, the $D$. The spatial diffusion coefficient is computed $D_x=T^2/D$. In this calculation, two different values of $Q_s$ are considered, $Q_s=1$ GeV and $Q_s=2$ GeV. It is important to mention that the estimation of the effective temperature in the pre-equilibrium phase involves uncertainty, hence, in the $D_x$.  The $D$ obtained in 
the pre-equilibrium phase is quite close to the value obtained within pQCD at the same temperature. As shown in Fig.~\ref{fig:Dx}, 
the $D_x$ evolves continuously from the pre-equilibrium phase to the QGP phase. The heavy quark thermalization time in the QGP phase can be obtained from the $D_x$ using equation ~\ref{eq:th}. The $D_x$ obtained within the transport approach are well within the uncertainty of lattice QCD results. This highlights the significance of the heavy quarks as a probe of QGP, which have the potential
to connect the phenomenology constrained by the experimental data to lQCD to study the
transport properties of the hot QCD matter. A recent study showed that the memory effect~\cite{Ruggieri:2022kxv} slows down heavy quark momentum evolution in QGP. This indicates one requires a large momentum diffusion coefficient, hence, smaller $D_x$, to reproduce the same $R_{AA}$.

\section{Recent developments}
Temperature dependence of the transport coefficients plays an important role for a simultaneous description of both the $R_{AA}$ and $v_2$~\cite{Das:2015ana}. Hence, the simultaneous study of both observables can constrain the temperature dependence of the heavy quark transport coefficients in QGP and disentangle different energy loss models. In a recent study, it has been shown that the event-by-event correlations~\cite{Plumari:2019hzp}, a measure of the linear correlation, is given by the correlation coefficient 
$C(n,m)$ as:
\begin{equation}
C(n,m)=\frac{\sum_{i}(v_n^{L,i}-\langle v_n^{L} \rangle)(v_m^{H,i}-\langle v_m^{H}\rangle)}{\sqrt{\sum_{i}(v_n^{L,i}-\langle v_n^{L} \rangle)^2\sum_{i}(v_m^{H,i}-\langle v_m^{H} \rangle)^2}},
\end{equation}

between light and heavy flavor flow harmonics, are novel observables to understand the heavy quark-bulk interaction and are sensitive to the temperature dependence  of heavy quark transport coefficients. In the above equation, $v_n^{L,i}$ and $v_m^{H,i}$ are the values of anisotropic flows corresponding to the event $i$ for light and heavy quarks respectively.  $C(n,m) \approx 1$ corresponds to a strong linear correlation between light and heavy quark anisotropic flows.

\begin{figure*}
\centering
\includegraphics[clip,height=5 cm]{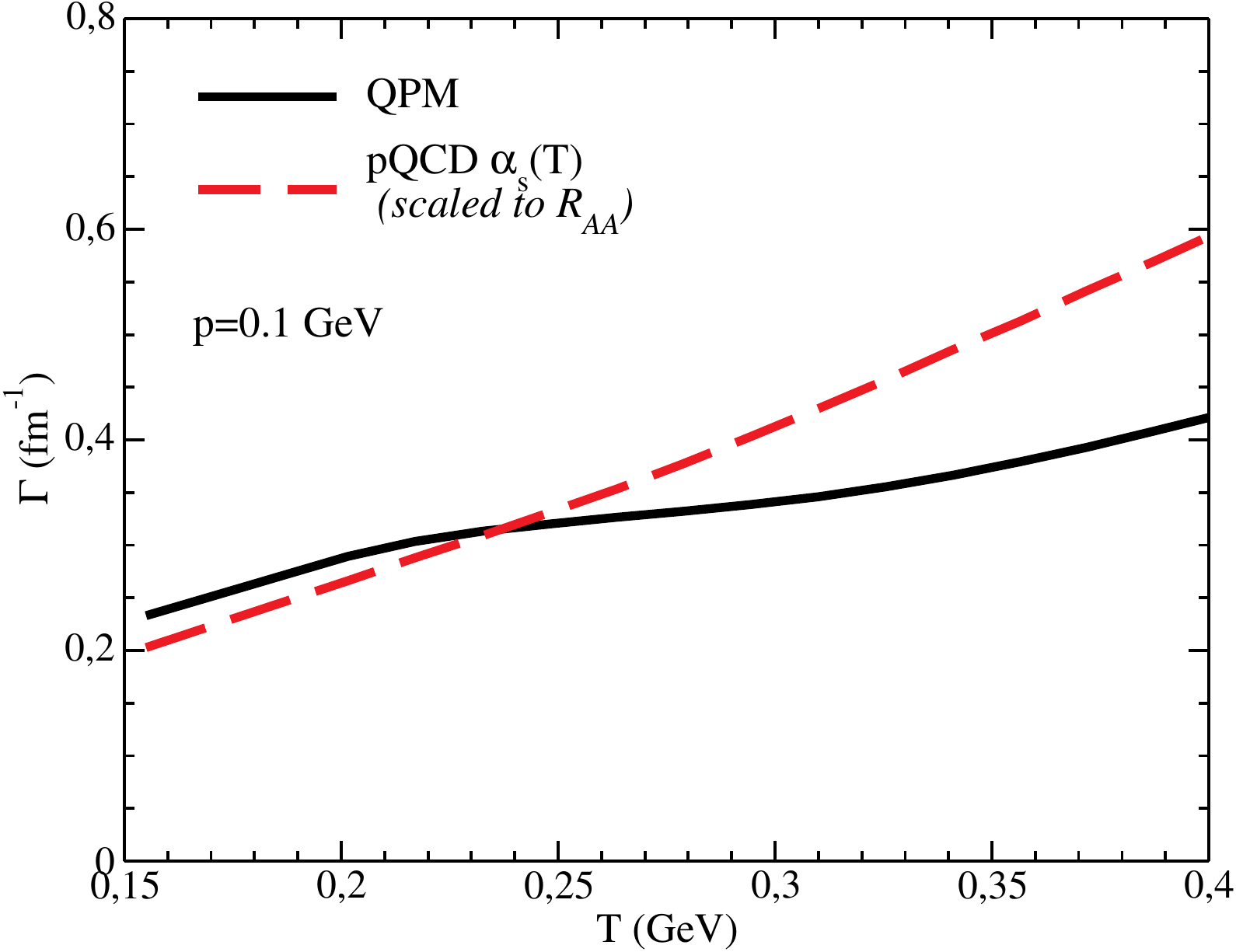}
\hspace{0.3cm}
\includegraphics[clip,height=5 cm]{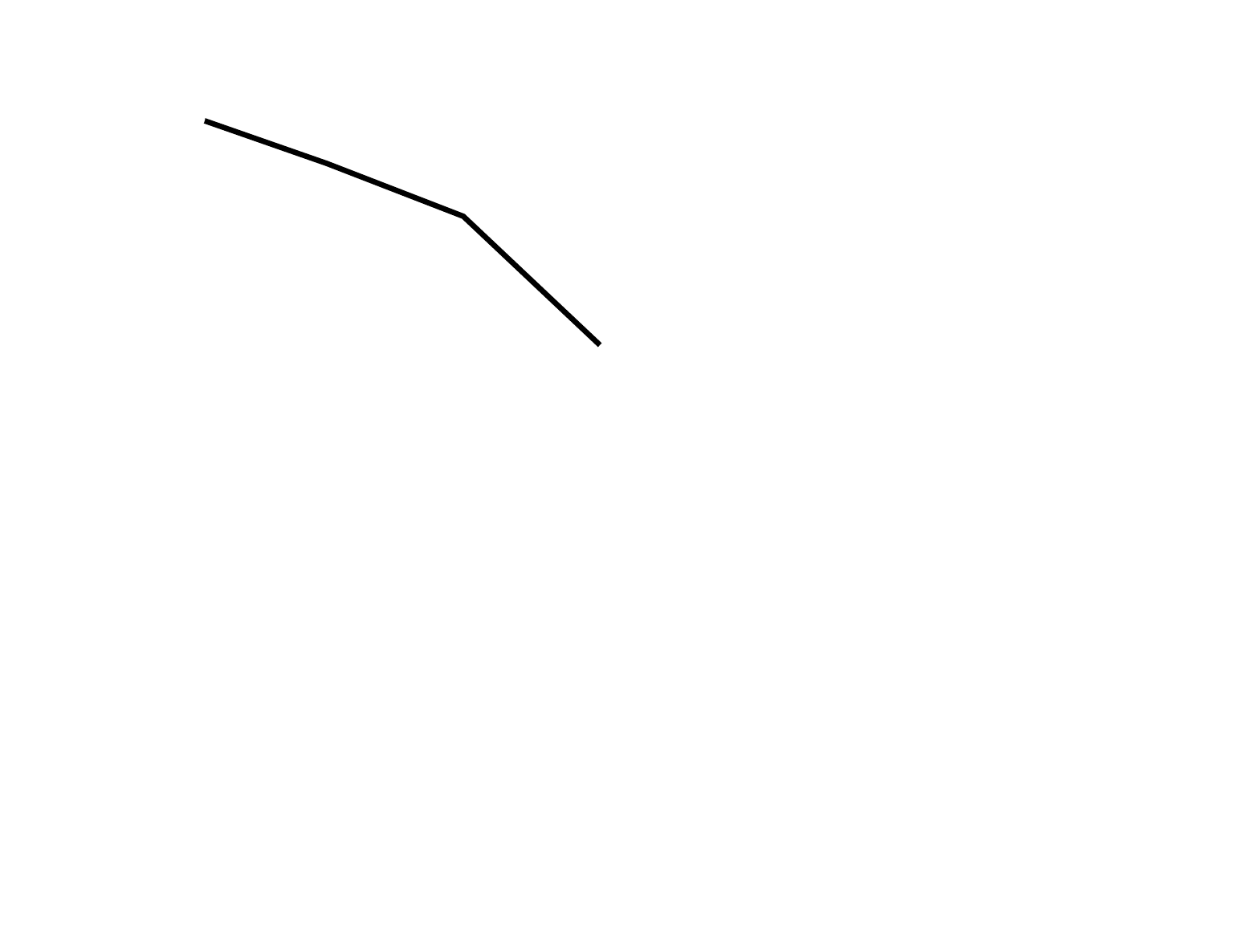}
\caption{Left panel: Temperature dependence of the drag coefficient $\Gamma$ obtained within QPM (black solid) and  pQCD (red dashed), scaled in order to reproduce the same $R_{AA}$~\cite{Plumari:2019hzp}. Right panel: Event-by-event correlation coefficient of D-meson $v_n$ and light hadron $v_n$  obtained within QPM (solid line) and pQCD (dashed line) as a function of 
the order of the harmonic $n$~\cite{Plumari:2019hzp}. } 
\label{fig:cor}
\end{figure*}

In Fig.~\ref{fig:cor} (left panel), the temperature dependence of the drag coefficient, $\Gamma$, obtained within QPM and pQCD is depicted which is rescaled  to reproduce the same $R_{AA}$. In Fig.~\ref{fig:cor} (right panel), the correlation coefficient $C(n,n)$ as a function of the order of the harmonic $n$ is shown within both QPM and pQCD to understand the impact of 
the temperature dependence of heavy quark transport coefficients is presented. As shown in Fig.~\ref{fig:cor} (right panel), the  correlation is stronger for QPM than pQCD and sensitive to the temperature dependence of the drag coefficient, $\Gamma$. 
If  this correlation coefficient is measured in the experiment, it can further constrain the temperature dependence of the heavy quark transport coefficients. 

Heavy quark transport coefficients are usually calculated assuming the
collision partners from the QGP are in thermal equilibrium. This assumption may not always
correct for heavy-ion collisions considering the fact that the bulk particles need time to reach  thermal equilibrium and 
particles with large momentum are far from thermal equilibrium. Hence, the non-equilibrium effects should be considered 
to compute the heavy quark transport coefficients. Non-equilibrium effects are introduced in ref.~\cite{Song:2019cqz} in different possible scenarios 
to compute the heavy quark transport coefficients. It is shown that each non-equilibrium scenario affects the charm 
quarks' transport coefficients in a different way and the modifications are about 20-30$\%$. 
Another interesting development is the investigation of the  nonperturbative effects of gluon emission off heavy quarks 
propagating through the QGP within the T-matrix~\cite{Liu:2020dlt}. They incorporate the nonperturbative effects step by step by including the confining interactions, resummation in the heavy-light scattering amplitude, and off-shell spectral functions for both heavy and light quarks. They observe that the perturbative processes encounter a strong suppression of soft radiation due
to the thermal masses of the emitted gluons. The nonperturbative effects enhance the radiative contribution at low 
momentum and temperature. However,  its magnitude is small compared to the elastic contribution.

To describe the heavy quark $R_{AA}$ and $v_2$  simultaneously at low momentum requires the nonperturbative contribution. 
Recently, the linear Boltzmann transport model has been extended by implementing a Cornell-type potential~\cite{Xing:2021xwc} inspired
by Ref.~\cite{Liu:2017qah} that incorporates both short-range Yukawa interaction and long-range color confining interaction for the 
heavy quark and QGP. They have extracted the in-medium heavy quark potential through the model-to-data comparison which is in reasonable agreement with the lattice QCD data. Considering both the collisional and radiative loss within the DREENA-A~\cite{Zigic:2021rku}, they recently incorporated 3+1D expansion  to compute both the $R_{AA}$ and $v_2$  for light and heavy flavored hadron which
can able to describe the data  at high $p_T$.

In a small system produced at p-Pb collisions, experimental measurements indicate a surprisingly large $v_2$ for the D meson~\cite{CMS:2018loe}.
However, the nuclear  suppression factor, $R_{pPb}$, is close to one. Simultaneous description of both the $R_{pPb}$ and $v_2$ 
in small system is a top challenge and very few attempts have been made so far~\cite{Beraudo:2015wsd,Zhang:2020ayy,Liu:2019lac}. Recently, it has been shown~\cite{Zhang:2022fum} that within AMPT 
considering both the parton interactions and  the large Cronin effect, one can able to describe both the observables; the Cronin effect enhances the charm quark yields and cancels out the effect of jet quenching~\cite{Zhang:2022fum}. However, they have incorporated a large constant cross-section for the heavy quark light parton interaction and large Cronin effect. System-size scan~\cite{Katz:2019qwv} of D meson $R_{AA}$ and $v_2$ 
can through light to understand the heavy quark dynamics in small systems. It has been shown that in central collisions the D meson $v_2$ is almost independent of system size. However, the $R_{AA}$ approach to one as the system size decreases. 
This is mainly due to the  interplay between the shrinking path length and the enhancement of eccentricities in small systems, due to which the $v_2$ remains almost unchanged at high multiplicity.

The heavy baryon to heavy meson ratios are fundamental for the understanding of the in-medium hadronization  of the heavy hadrons 
 with respect to the light flavored baryon to meson ratio~\cite{Ghosh:2014oia}. Both RHIC and LHC measured a large $\Lambda_c/D$ ratio~\cite{STAR:2018zdy,ALICE:2020wfu} in nucleus-nucleus collisions. Enhancement of $\Lambda_c/D$ ratio will affect the D meson nuclear modification factor, $R_{AA}$. Heavy baryon to heavy meson ratios can serve as a tool to disentangle different hadronization models as it is very sensitive to harmonization mechanisms~\cite{Das:2016llg,Plumari:2017ntm}. Understanding 
heavy baryon to heavy meson ratios at different colliding systems is a top challenge.  In ref.~\cite{He:2019vgs},  a 4-momentum conserving recombination model for baryons is developed, which includes space-momentum correlations between charm quarks and the hydro medium and maintains HQ number conservation. In this model, the equilibrium limit is improved by incorporating a large set of “missing” heavy flavor baryon states not listed by the particle data group but predicted by the relativistic-quark model(RQM), consistent with lQCD calculations. Recently the Catania group developed a coalescence plus fragmentation model for charm quark hadronization at LHC energy at pp collisions~\cite{Minissale:2020bif} (see also ~\cite{He:2019tik}). The model is essentially the same as the one developed for nucleus-nucleus collisions without jet quenching. Within this model, they can able to describe $\Lambda_{c}^{+}/D^{0}$, $D_s^+/D^0$ and $\Xi_{c}^{0,+}/D^0$ at pp collisions at LHC energy. For more details, we refer to ref. ~\cite{Beraudo:2023fpq}.

\section{Heavy quark as a probe of the initial stage effects}
Till recently, heavy quark momentum evolution in the pre-equilibrium phase~\cite{Das:2017dsh}, before the formation of QGP, was approximated 
by free streaming. In the pre-equilibrium phase, the energy density is very high; this early stage dynamics can be quite relevant, especially for heavy quarks considering their short formation time~\cite{Mrowczynski:2017kso,Ruggieri:2018rzi,Sun:2019fud,Liu:2020cpj,Carrington:2020sww,Boguslavski:2020tqz,Boguslavski:2023fdm,Avramescu:2023qvv}. Within the color glass condensate (CGC) effective theory the pre-equilibrium stage of the high energy collisions can be described in terms of strong gluon fields, namely the Glasma. 
Recently it has been shown that these fields induce strong diffusion of charm quarks in momentum space~\cite{Ruggieri:2018rzi,Sun:2019fud,Liu:2020cpj}. As a consequence, low $p_T$ charm quarks shifted to high $p_T$, and  the spectrum of charm quarks tilted towards higher $p_T$. 
The tilting of the spectrum affects the nuclear suppression factor, the ratio between the final and the initial D meson spectrum.

\begin{figure*}
\centering
\includegraphics[clip,height=5.5 cm]{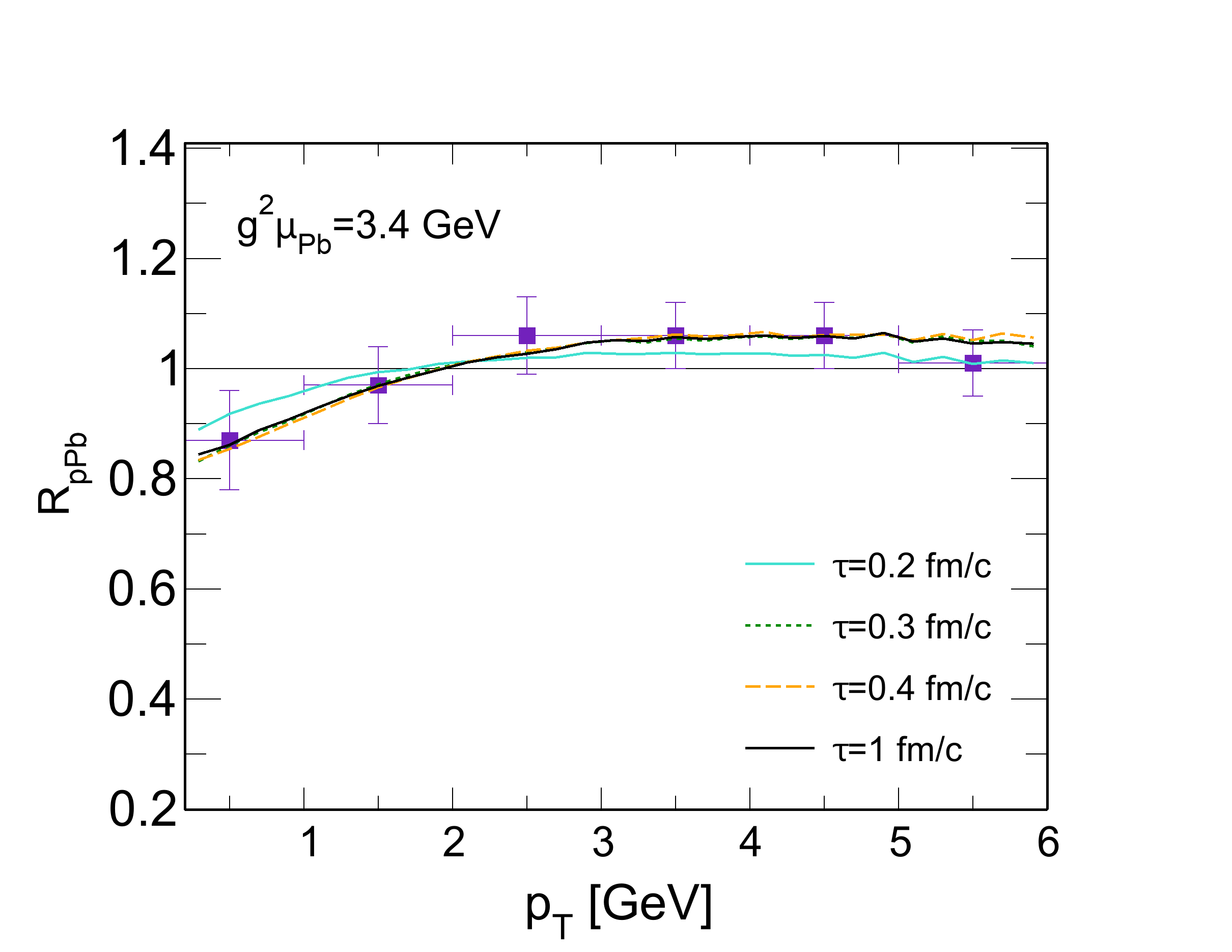}
\hspace{0.3cm}
\includegraphics[clip,height=5.5 cm]{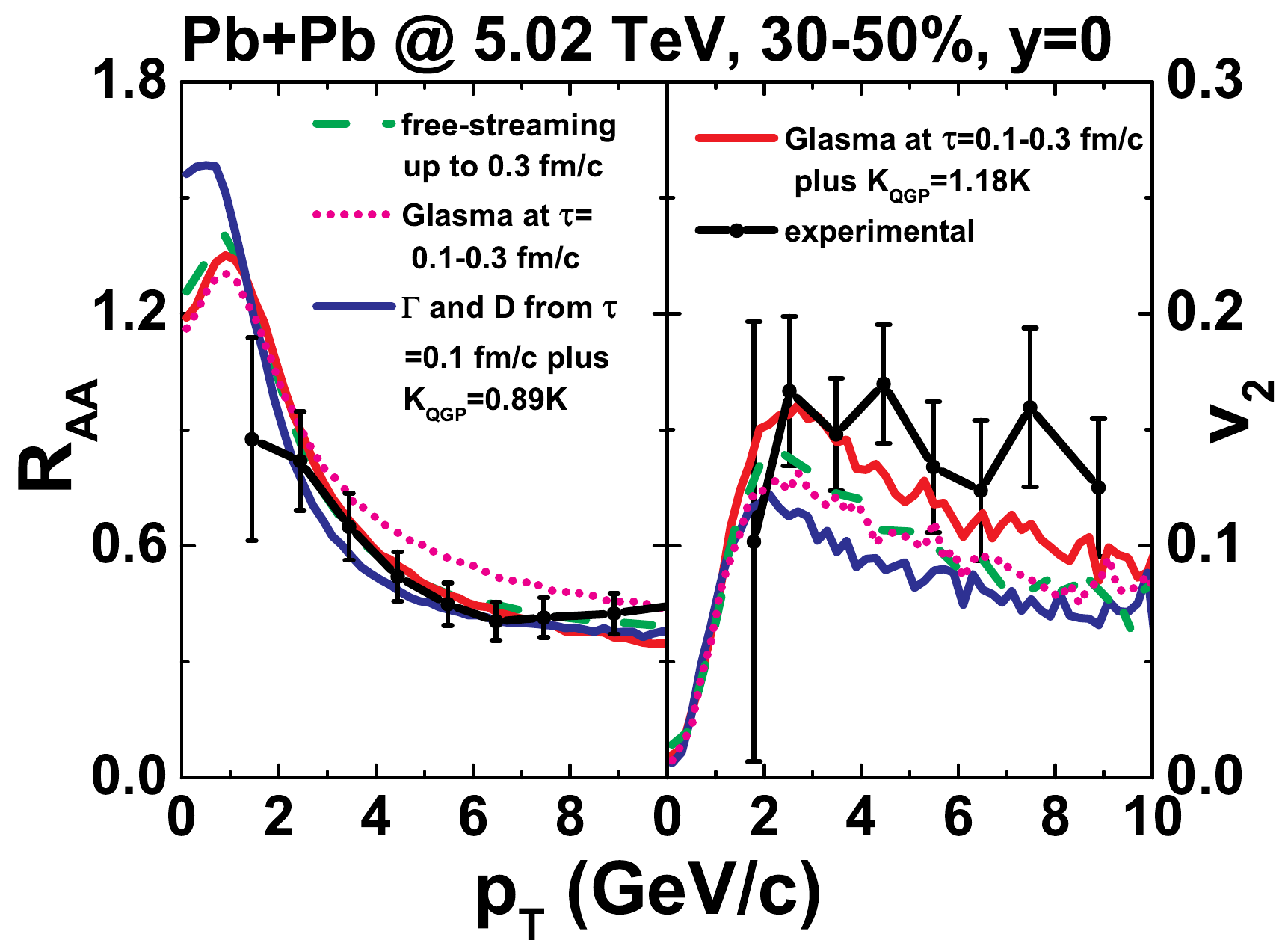}
\caption{Left panel: Nuclear suppression factor  for the $D^0$-meson versus $p_T$ at different times in p-Pb collisions at $\sqrt{s}\!=\!5$ TeV ~\cite{Liu:2019lac}. Right panel: $R_{AA}$  and $v_2$ of D mesons for different evolutions with and without Glasma phase~\cite{Sun:2019fud}:
(a) free streaming up to $0.3\, \rm fm/c$ followed by drag and diffusion in the QGP phase (green dashed line); 
(b) with Glasma dynamics up to $0.3\, \rm fm/c$ and then evolution as in (a) shown by magenta dotted line; 
c) with the drag and diffusion dynamics starting at $\tau_0= 0.1 \,\rm fm/c$ shown by blue solid line with K factor decreasing by 11\% relative to (a); (d) with Glasma dynamics up to $0.3\, \rm fm/c$ and then evolution with K factor increasing by 18\% relative to (a) shown by red solid line } 
\label{fig:pp-RAAv2}
\end{figure*}

In Fig.~\ref{fig:pp-RAAv2} (left panel), the D meson nuclear suppression factor, $R_{pPb}$,  is shown at different times for D meson with $g^2\mu=3.4$ GeV for the Pb nucleus. The momentum evolution of charm quarks in Glasma has been studied via the Wong equations; then at the time shown in the figure,  the fragmentation procedure is adopted to get the $D-$meson spectra.
The $R_{pPb}$ substantial  deviation  from one is because of the interaction of the charm with the gluon fields produced in the 
p-nucleus collisions. The shape of the $R_{pPb}$ computed is in qualitative agreement with the experimental data at the same energy~\cite{Liu:2019lac}. In the p-nucleus collisions,  the impact of the QGP phase is ignored. However, in nucleus-nucleus collisions, the Glasma 
phase can be considered as the initial condition for the QGP phase. In Fig.~\ref{fig:pp-RAAv2} (right panel), the D meson $R_{PpPb}$ and $v_2$ are compared with the experimental results with and without the Glasma phase for different possible initializations~\cite{Sun:2019fud}. It is important to mention in the Glasma phase the $R_{PpPb}$ will get  enhanced at intermediate $p_T$. To reproduce the same $R_{PpPb}$ with the Glasma phase and without the Glasma phase, one requires to enhance the interaction with the Glasma phase to compensate for the initial enhancement of the $R_{PpPb}$. As a consequence, it will enhance the $v_2$ with the Glasma phase. Hence, the inclusion of the Glasma phase will help for a simultaneous description of both the $R_{PpPb}$ and $v_2$. This initial Glasma phase can play a significant role in the small system produced in p-nucleus collisions~\cite{Sun:2023adv}. Recently, the heavy quark diffusion coefficient has been evaluated in ref.~\cite{Boguslavski:2020tqz} in an over-occupied gluon plasma. They have observed  a rapid initial rise in the momentum broadening. It is important to note that Glasma phase dynamics can not be mimicked within Langevin dynamics considering the fact that in the Glasma phase, heavy quarks experience diffusion without substantial drag.

Heavy quarks are also considered as an excellent probe of the produced electromagnetic fields and  initial tilt of the fireball produced in non-central heavy-ion collisions which can be characterized through the directed flow $v_1$ of the D-meson~\cite{Das:2016cwd}. The electromagnetically-induced splitting, $\Delta v_1$, in the directed flow of charm and anti-charm through  $D$ and  $\overline{D}$ meson can characterize the produced electromagnetic field. Recently, both the STAR and ALICE collaborations~\cite{STAR:2019clv,ALICE:2019sgg} have measured the $v_1$ and $\Delta v_1$ of the D meson and observed a non-zero $v_1$ of the D meson. 

\begin{figure*}
\centering
\includegraphics[clip,height=5 cm]{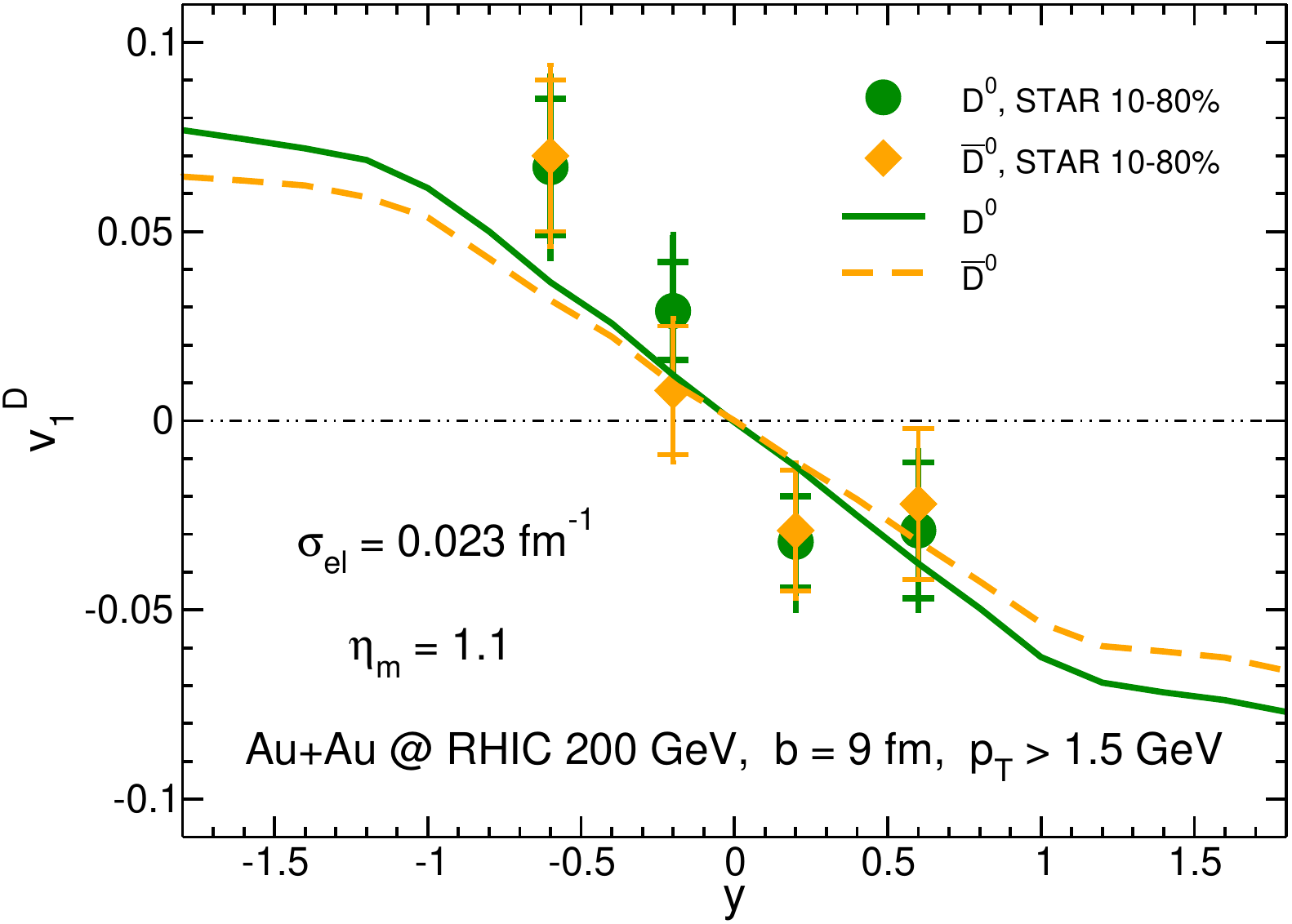}
\hspace{0.3cm}
\includegraphics[clip,height=5 cm]{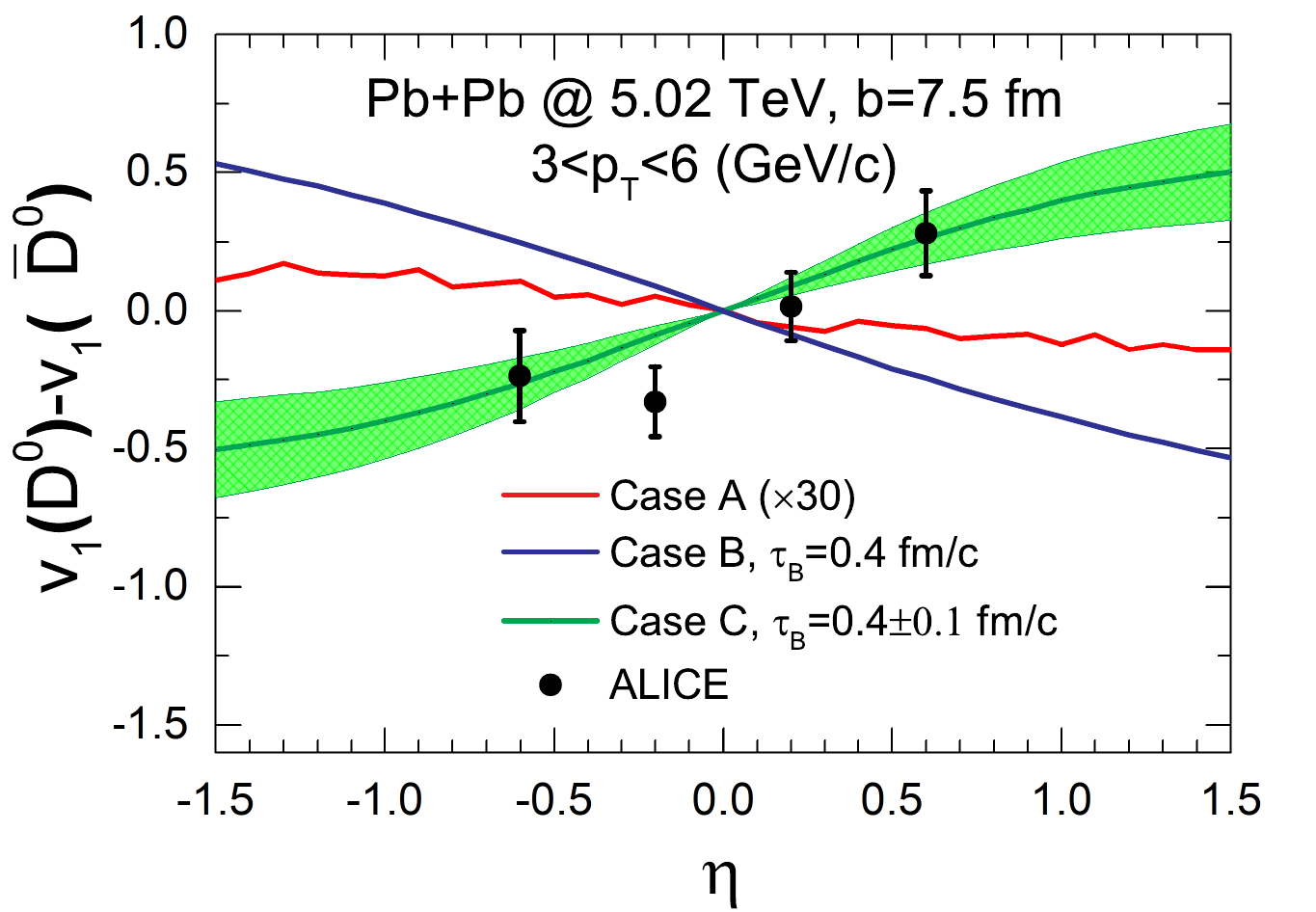}
\caption{Left panel:Directed flow of $D$ and $\overline{D}$ mesons~\cite{Oliva:2020doe} as a function of rapidity for Au+Au collisions at RHIC energy $\sqrt{\sigma_{NN}}=200$ GeV in comparision with  experimental data from STAR Collaboration~\cite{STAR:2019clv}. Right panel:$v_1(D^0)-v_1(\overline{D}^0)$ ~\cite{Sun:2020wkg} as a function of pseudorapidity for three different cases for Pb+Pb collisions and compared to experimental data from  ALICE collaboration~\cite{ALICE:2019sgg}.  } 
\label{fig:v1}
\end{figure*}

 The $D^0$ meson $dv_1/dy$ measured at the highest RHIC energy by STAR Collaboration is about 25 times larger than that of the charged kaons. The $\Delta v_1$ of D mesons at the  highest RHIC energy is smaller than the current precision of the measurement. The ALICE Collaboration at LHC  reported a positive slope for the $\Delta v_1$,  which is about  3 orders of magnitude larger than  that of  the light-charged hadrons. Very few attempts have been made so far 
to compute the D meson directed flow~\cite{Das:2016cwd,Chatterjee:2018lsx,Oliva:2020doe, Oliva:2020mfr, Dubla:2020bdz, Sun:2020wkg}  both at RHIC and LHC energies. All the model calculations reported a finite directed flow for the D meson. However, they obtained a negative slope for the D-meson $\Delta v_1$. As shown in ref.~\cite{Sun:2020wkg}, in a phenomenological approach,  one can reproduce a positive slope of the $\Delta v_1$ only if the magnetic field dominates over the electric field. They indicate that the time evolution of the electromagnetic field plays an important role in determining the slope of the $\Delta v_1$.

\section{Summary}
The study of heavy quark dynamics in QCD matter is quite complex and
requires considering its propagation through the pre-equilibrium phase, QGP phase, and hadronic phase after hadronization to compare with the experimental observables. Over the years since the first experimental data, several advancements have been made to understand the heavy quark dynamics in QCD matter and to extract the heavy quark transport coefficients. Non-perturbative effects along with hadronization through coalescence plus fragmentation is the key ingredient to describe both the $R_{AA}$ and $v_2$ simultaneously.
Present calculations indicate heavy quark thermalization time is about 2-6 fm/c at low $p_T$.  

Since recently we used to ignore the role of the pre-equilibrium phase considering its lifetime is very short. However, recent calculation indicates that the pre-equilibrium phase is playing a significant role and can alter the $R_{AA}$ and $v_2$ dynamics for a simultaneous description of both the observables in nucleus-nucleus collisions. The pre-equilibrium phase can play an important role in p-nucleus collisions 
considering the fact that the lifetime of the QGP phase might be very small in p-nucleus collisions. Recent calculations indicate that with the pre-equilibrium phase, the Glasma phase, one can reproduce the experimental data on $R_{pPb}$.  It is important to note that Glasma phase dynamics can not be mimicked within Langevin dynamics considering the fact that in the Glasma phase, heavy quarks experience strong diffusion without substantial drag. Heavy quark directed flow is a potential observable to characterize the strong electromagnetic fields produced in high-energy collisions.  
Time evolution of the electromagnetic field plays a crucial role in
determining the slope of the $\Delta v_1$. For a complete study, it will be important to include the effect of electromagnetic fields on  the heavy quark transport coefficients~\cite{Fukushima:2015wck, Kurian:2019nna} while evaluating the $v_1$.

So far the major focus is to study heavy quark dynamics at RHIC and LHC energies. However, it's time to focus on heavy quark dynamics at small systems produced in p-nucleus collisions and  beam energy scan. Considering the fact  that the momentum evolution 
of the bottom quarks is approximately that of Brownian motion~\cite{Das:2013kea}, the bottom quark can act as a novel probe to understand the transport properties of hot QCD matter. New observables like heavy light even-by-event correlation and $D-\overline{D}$ correlation can further constrain the temperature dependence of the heavy quark transport coefficients and disentangle different energy loss mechanisms.

Acknowledgments: SKD acknowledges valuable discussions with Marco Ruggieri and Pooja.

\end{document}